\documentclass[10pt,journal,compsoc]{IEEEtran}

\ifCLASSOPTIONcompsoc
  \usepackage[nocompress]{cite}
 \usepackage[caption=false,font=footnotesize,labelfont=sf,textfont=sf]{subfig}
\else
 \usepackage[caption=false,font=footnotesize]{subfig}
  \usepackage{cite}
\fi
\ifCLASSOPTIONcaptionsoff
 \usepackage[nomarkers]{endfloat}
\let\MYoriglatexcaption\caption
\renewcommand{\caption}[2][\relax]{\MYoriglatexcaption[#2]{#2}}
\fi
\usepackage{url}
\usepackage{fixltx2e}
\usepackage{enumitem}
\usepackage{array}
\usepackage{algorithmic}
\usepackage{amsmath}
\usepackage{graphicx}
\usepackage{tabularx}
\usepackage{multirow}
\usepackage{moreverb}
\usepackage{rotating}
\usepackage{tcolorbox}
\usepackage{dialogue}
\usepackage{float}
\usepackage{hyperref}
\usepackage{cleveref}
\usepackage{footnote}
\usepackage{mathrsfs}
\usepackage{booktabs}
\usepackage[htt]{hyphenat}
\usepackage{dblfloatfix}
\usepackage{wrapfig}
\usepackage{fancyvrb}
\usepackage{ragged2e}
\usepackage{multicol}

\newfloat{excerpt}{h}{lob}
\floatname{excerpt}{Excerpt}
\Crefname{excerpt}{Excerpt}{Excerpts}
\newcolumntype{Y}{>{\centering\arraybackslash}X}

\begin{document}

\title{Programmer Visual Attention During Context-Aware Code Summarization}

\author{Robert Wallace, Aakash Bansal, Zachary Karas, Ningzhi Tang, Yu Huang, Toby Jia-Jun Li, Collin McMillan
\IEEEcompsocitemizethanks{
  \IEEEcompsocthanksitem Manuscript received ---- ---- -----. This work is supported in part by the NSF CCF-2100035. Any opinions, findings, and conclusions expressed herein are the authors’ and do not necessarily reflect those of the sponsors.
  \IEEEcompsocthanksitem This work involved human participants in its research. Approval of all ethical and experimental procedures and protocols was granted by Internal Review Boards of University of Notre Dame under Application No. 22-03-7128 and Vanderbilt University under No. 220604.
  \IEEEcompsocthanksitem The authors are with the Department
  of Computer Science and Engineering, University of Notre Dame, IN 46556, USA.\protect\\
  E-mail: \{rwallac1,abansal1,ntang,toby.j.li,cmc\} @nd.edu\\
  and with the Department of Computer Science, Vanderbilt University, TN 37235, USA.\protect\\
  E-mail: \{z.karas,yu.huang\} @vanderbilt.edu
  \IEEEcompsocthanksitem This paper has supplementary downloadable multimedia material available
  at 
\url{https://github.com/aakashba/EyeContext-TSE} provided by the authors.
}}

\IEEEtitleabstractindextext{%
\begin{abstract}
Programmer attention represents the visual focus of programmers on parts of the source code in pursuit of programming tasks. The focus of current research in modeling this programmer attention has been on using mouse cursors, keystrokes, or eye tracking equipment to map areas in a snippet of code. These approaches have traditionally only mapped attention for a single method. However, there is a knowledge gap in the literature because programming tasks such as source code summarization require programmers to use contextual knowledge that can only be found in other parts of the project, not only in a single method. To address this knowledge gap, we conducted an in-depth human study with 10 Java programmers, where each programmer generated summaries for 40 methods from five large Java projects over five one-hour sessions. We used eye tracking equipment to map the visual attention of programmers while they wrote the summaries. We also rate the quality of each summary. We found eye-gaze patterns and metrics that define common behaviors between programmer attention during context-aware code summarization. Specifically, we found that programmers need to read up to 35\% fewer words (p\textless0.01) over the whole session, and revisit 13\% fewer words (p\textless0.03) as they summarize each method during a session, while maintaining the quality of summaries. We also found that the amount of source code a participant looks at correlates with a higher quality summary, but this trend follows a bell-shaped curve, such that after a threshold reading more source code leads to a significant decrease (p\textless0.01) in the quality of summaries. We also gathered insight into the type of methods in the project that provide the most contextual information for code summarization based on programmer attention. Specifically, we observed that programmers spent a majority of their time looking at methods inside the same class as the target method to be summarized. Surprisingly, we found that programmers spent significantly less time looking at methods in the call graph of the target method. We discuss how our empirical observations may aid future studies towards modeling programmer attention and improving context-aware automatic source code summarization.
\end{abstract}

}

\maketitle

\section{Introduction}

Code summarization is the act of writing a natural language description for a snippet of source code, such as a subroutine. These summaries are meant to fill the knowledge gap between a higher-level understanding of the program that is necessary to use it, and the lower-level implementation in the code. The automatic generation of these summaries has been an important goal for the software engineering research community for decades. Lately, tools such as GitHub Copilot and OpenAI ChatGPT with automatic code summarization capabilities are increasingly becoming part of Software Engineering workflows~\cite{camara2023assessment}. At the core of these tools are Large Language Models (LLMs) that learn representations of both the programming language and natural language in an attempt to fill the aforementioned knowledge gap. They can be thought of as \emph{translating} code into natural language summaries.

However, these summaries do not necessarily reflect real-world scenario, as they are not ``context-aware''. A summary is context-aware when it considers the context, i.e., other code in the project related the subroutine either required to run it or understand it.~\cite{McBurney:ICPC:2014,krinke2006effects} Although there are several types of context, for the purposes of this study, context is source code from the project related to the subroutine. Context-aware summaries are important because a subroutine usually does not exist in a vacuum; the manner in which a subroutine is used depends on the context. Therefore project-level information in critical to writing a context-aware summary. Over the last five years, a few studies have introduced context to code summarization models, such as by adding source code from part of the project ~\cite{haque2020improved,bansal2021project} or the call-graph~\cite{bansal2023callcon}. It is cost-prohibitive to feed the entire project to these language models as a longer input requires more computational resources. There is a knowledge gap in related work on how human programmers navigate through the context in a project to infer the most important information. That is the core inspiration behind this study. We designed this study to discover what programmers’ visual attention is during code summarization and to discover what contextual information they find most important, so that we may use that information for future studies in automated source code summarization.

The visual attention of a programmer is an important indication of their mental model when performing programming tasks~\cite{bednarik2012expertise,Rodeghero:ICSE:2014,sharafi2015systematic}. Visual attention strategies refer to the selective focus humans apply to some parts of the visual stimulus (i.e., the code in the case of programmers). This selection of parts of the code is informed by the mental model of the programmer that evolves as they retain more information ~\cite{bednarik2008temporal}. These visual attention strategies can be inferred using eye tracking data, specifically, gaze patterns of programmers. eye trackers have been used in fields such as computer vision, psychology, and medical sciences to create models of human behavior and mental processes for decades~\cite{ouerhani2004empirical,hood1998adult,olivers2006feature}. Although models of visual attention strategies are interesting on their own, they can used to learn 
domain-specific knowledge such as those programmers use while completing SE tasks.

To that end, recent work in SE towards modeling programmer attention has mainly used proxies such as mouse cursor movements and clicks~\cite{paltenghi2021thinking}. Studies to model visual attention directly using eye trackers have been limited to a small snippet of code~\cite{bednarik2006eye, Rodeghero:ICSE:2014, karas2024tale}. Recently, a few studies have also proposed approaches for automatic inference of visual attention~\cite{bansal2023human,bansal2023scanpath,zhang2024eyetrans}. The scope of most of these studies is limited to one subroutine. We build upon these works to study eye tracking and visual attention patterns at the project-level. Compared to those studies, our study design is challenging due to difficulty controlling a study environment where the participants have access to the entire project. Analyzing the data at the project level also poses challenges as not all code in the project is useful or even executed. However, project-level comprehension is a real world scenario in which programmers complete SE tasks such as code summarization. To the best of our knowledge, we conducted the first comprehensive human study to map these attention patterns in the context of the project.

In this paper, we present an eye tracking study designed with the goal of analyzing visual attention strategies of programmers while they navigate related methods for the task of source code summarization. We recruited 10 Java programmers, each tasked with completing five sessions. In each session, we asked programmers to write context-aware summaries for 8 subroutines of a Java project. Then, we manually rated the summaries for accuracy, completeness, conciseness, and clarity using two graders, who are among the authors. These ratings help us analyze the effect of various programmer gaze patterns on the quality of summaries. Finally, we categorized all the code in the project into categories of code context. We then identified the type of context to which the programmers gave the highest attention in pursuit of the summarization tasks. Although knowledge of programmer visual attention strategies is intellectually interesting, we posit that models of programmer attention and identification of important parts of context have practical applications towards improving models for automatic source code summarization.

We found that: 1) participants looked at fewer words and revisited words fewer times as they processed more methods for a given project; 2) the quality of these summaries did not suffer or decrease with any statistical significance as the participants processed more tasks; 3) the quality of summaries is related to the amount of context analyzed by the participants, like a bell-curve, such that the quality increases as more context is analyzed by the participants up to a certain threshold and then the quality decreases; 4) participants spent the most time looking at class and instance methods, followed by class declarations and call graphs; 5) the type of context participants focused on--such as class and instance methods, declarations, the call graph, or other areas of the project--is generally not affected by the project, session order, method order, or participant. 

We make five contributions in this study:
\begin{enumerate}
    \item We perform an eye tracking study involving more than 60 hours
    of participant effort by 10 Java programmers. Each programmer completed 40 context-aware code summarization tasks. 
    \item We contribute a novel dataset of 394 context-aware summaries, to be released publicly with this study. 
    \item We perform a qualitative comparison of the context-aware summaries, graded by two authors of this paper and agreed upon by a discussion-based coding methodology, to be released publicly with this study.
    \item We perform a quantitative comparison of eye-gaze patterns and visual attention strategies as the programmers process more tasks and their correlation with quality of summaries.
    \item We present a quantitative comparison of the varying levels of attention received by different types of context and make recommendations to distill context for future automatic source code summarization studies.
\end{enumerate}

\section{Background and Related Work}
In this section, we provide background on eye tracking studies in software engineering and discuss related work in context-aware code summarization.

\vspace{-0.3cm}
\subsection{Eye tracking in Software Engineering}

Eye tracking has been used in SE research for nearly two decades to gain insight into programmer behavior and human factors in programming~\cite{sharafi2015systematic}. Recent evolution of eye trackers from expensive and difficult-to-operate lab equipment to simple monitor-mounted devices usable in development environment has accelerated their use in SE research~\cite{braw2023integrating,bansal2023scanpath}. Eye tracking studies in SE can mainly be divided into three areas of interest:

\textbf{Code Comprehension} is an area of interest that is primarily concerned with the mapping of programmer mental model during comprehension of a snippet of code or small program. In 2006, Aschwanden and Crosby~\cite{aschwanden2006code} presented gaze patterns of programmers during the comprehension of a small snippet of code containing a loop and mathematical equations. They formalized the process of understanding programmer mental processes using eye tracking during program comprehension. Also in 2006, Bednarik and Tukiainen~\cite{bednarik2006eye} introduced a framework for analyzing eye gaze patterns, in particular scanpaths, of participants tasked with comprehension of two small programs. They found that the mental model of a programmer is refined as they progress through the task, marked by a decrease in attention switching. In 2014, Rodeghero~\emph{et al.}~\cite{rodeghero2014improving} presented an analysis of gaze patterns of programmers tasked with summarization of source code. They used the findings of that study to improve automatic source code summarization, showing how code summarization approaches of the time could be improved by making them focus on the same things as people do. Abid~\emph{et al.}~\cite{abid2019developer} replicated and expanded the Rodeghero study~\cite{rodeghero2014improving} in 2019 with project context using a similar interface, and showed that allowing the participants to work in a more realistic environment results in a more generalized understanding of developer behavior. 

Similarly, in 2024, Karas~\emph{et al.}~\cite{karas2024tale} presented and compared visual attention strategies of programmers for two tasks: reading of source code summaries for assessment and generating their own source code summaries. They found that programmers tend to look at specific words in the method in specific order defined by the syntactic nature of the code. Their work serves as an inspiration for this study, however    these studies only considered method-level information. In this study, we focus our analysis on the project-level information.

\textbf{UML Diagrams} are an area of interest concerned with understanding how programmers use UML diagrams during programming tasks. In 2006, Guéhéneuc~\cite{gueheneuc2006taupe} presented two case studies on the importance of UML diagrams in program comprehension by following the gaze patterns of programmers presented with a comprehension task. In 2010, Sharif and Maletic~\cite{Sharif:2010:ELD:1796177.1796642} showed that eye gaze patterns of programmers suggest better program comprehension when presented with particular UML layouts and clustering patterns. A caveat of studies on UML diagrams is that the results are specific to the visualization tools and programmers were not presented source code or asked to do a programming task~\cite{lubke2021influence,Sharafi2020PracticalGuide}. Therefore, the observations from these studies do not necessarily represent visual attention during code comprehension.

\textbf{Code Debugging} studies are concerned with mapping programmer attention and mental model during the task of code debugging~\cite{sharafi2015systematic,Obaidellah2018}. In 2008, Bednarik and Tukiainen~\cite{bednarik2008temporal} presented an analysis of gaze patterns of programmers with varying levels of experience during debugging tasks. They found that more experienced programmers tend to use context-switching quicker and more often that novice programmers. In 2014, Turner~\emph{et al.}~\cite{turner2014eye} presented results of a study performed exclusively with students given two tasks: 1) over-viewing code to answer questions, and 2) debugging code. They found that gaze patterns are significantly different between the two tasks, suggesting eye-patterns of code comprehension are goal-dependent. In 2020, Sharafi~\emph{et al.}~\cite{sharafi2020eyes} presented a study that quantifiably indicates that developers generally follow a three-phase model of finding, learning, and editing. They found that developers spend more time searching code towards the beginning of the task, while what they focused on changed over time as they understood the program context. This is important because it highlights the need for in-depth studies like ours that specifically ask the programmer to summarize code using context, as well as finding out what parts of the code are more important to the programmer over time.

Our study is most closely related to code comprehension as we present participants with raw source code and specifically task them with summarization using context. The novelty of our study lies with the project-level eye tracking and data analysis.

\vspace{0.6cm}
\subsection{Source Code Summarization} 

Source code summarization has been a focus of SE research for a decade and a half. Early work in source code summarization used Information-Retrieval techniques~\cite{haiduc2010supporting,sridhara2011automatically}. 
Around 2015, these techniques were aided by the addition of contextual information~\cite{mcburney2016automated,zhang2016towards}. Around 2017, the advent of neural networks changed the landscape of automatic source code summarization research~\cite{iyer2016summarizing,hu2018summarizing}, but the initial approaches lacked contextual information.

In 2020, Haque~\emph{et al.}~\cite{haque2020improved} proposed an encoder to represent a few 
subroutines from the same file as the target subroutine. They showed that the addition of a small part of the file improved existing approaches in automatic source code summarization. 
Then in 2021, Bansal~\emph{et al.}~\cite{bansal2021project} extended upon that work by introducing a proof-of-concept project encoder that represents a few files from the project. 
They showed further improvement over file context by including more information but also cautioned against the costs of adding more context. 
They selected the files and subroutines in the files randomly and highlighted the need for a better methodology to narrow down contextual information. 

In 2023, Bansal~\emph{et al.}~\cite{bansal2023callcon} introduced a Graph Neural Network (GNN) based encoder to represent the call graph extracted from the project centered at the target subroutine. Recently, two studies used human visual attention to improve automatic source code summarization by using synthesized human attention data given limited eye tracking data~\cite{bansal2023human,zhang2024eyetrans}. However, the short summaries generated by these approaches may not represent the full potential of contextual information. Therefore, in this study we ask participants to write longer summaries that include specific contextual information in them.

Recent studies in automatic source code summarization have been focused on fine-tuning LLMs~\cite{su2024distilled,geng2024large,su2023language}. However, even for these massive models, the size of the input, called a ``context window'' is a limitation. Although there are proposed techniques for increasing the size of these context windows~\cite{zhang2024soaring}, there is also evidence that simple feeding a large amount of context 
to an LLM may not be helpful~\cite{bansal2023revisiting}. Therefore, in this study we aim to isolate specific areas in the project that programmers pay the most attention to, and may be used to distill contextual information for LLMs and any future techniques for automatic source code summarization.

\section{Study Design}
\label{sec:study}
In this section, we describe the experimental setup for our eye tracking study including the study setting, research questions, interface, procedure, dataset, and tools.

\begin{figure*}[t!]
	\centering
	\includegraphics[width=0.97\linewidth]{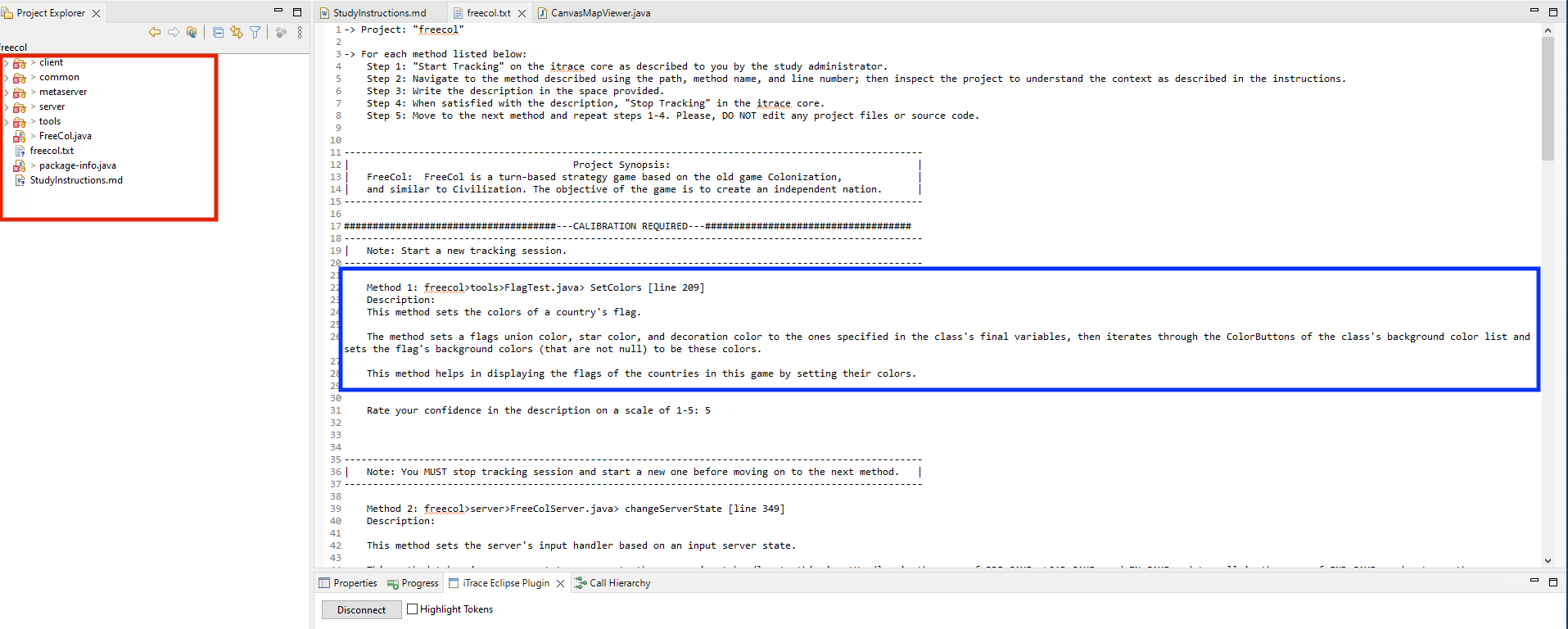}
    \vspace{-0.3cm}
	\caption{A screenshot of our interface. The blue box highlights the summary generated by the participants, while the red box highlights the navigational window limited to one project. The participant is free to open and read any part of the project.}
	\label{fig:interface}
    \vspace{-0.4cm}
\end{figure*}

\subsection{Study Scenario}
We design this study to emulate a specific scenario in code comprehension, specifically context-aware source code summarization. We ask participants to retrospectively write context-aware descriptions of a target method. We provide participants access to the entire Java code from the project that contains the target method. The project is devoid of any documentation, including header and inline comments (See \ref{sec:javaproj} on details of the projects). We ask the programmer to read and understand the target method, as well as the surrounding code in the context, and to write a three-sentence context-aware summary with the following instructions:
\begin{itemize}
\item The first sentence should explain the purpose of the method in as simple of terms as possible, in the context of the whole program.
\item The second should describe the more specific functionalities and under what circumstances they occur.
\item The last sentence should describe why this method is needed within the context of the overall project.
\end{itemize} 

We ask each programmer to write summaries for 8 methods per project, found to be the ideal number for 1 hour session during our pilot studies. We choose 5 projects extracted from Github for this study, and displayed only the raw Java code. We ask every programmer to process each project in a separate eye tracking session. We set up the study in the Eclipse IDE.

\vspace{-2mm}
\subsection{Research Questions}
We answer three research questions:

\begin{description}
\item[\textbf{RQ1:}] How do gaze patterns change as the participants progresses from the first method to the last in a session?

\item[\textbf{RQ2:}] How does the quality of summaries correlate with the gaze patterns?

\item[\textbf{RQ3:}] What parts of the project context received the highest attention? 
\end{description}

The rationale behind RQ1 is to evaluate how gaze patterns change as a participant moves from the first method to the last in a project. As participants investigate more methods and write descriptions, their knowledge base of the project may increase. We may be able to isolate specific gaze patterns associated with an increased knowledge base. 

The rationale behind RQ2 is to find gaze patterns that are characteristics of the quality of summaries that we grade manually. We aim to find the correlation of high-quality summaries and the observations from RQ1. We also aim to isolate gaze patterns that may be characteristic of writing what we consider to be higher-quality summaries.

The rationale behind RQ3 is that some parts of the context are more important to the task than others. We aim to categorize the different parts of context that the participants looked at and identify the type of context on which the participants spent the longest time fixating. This may help future studies distill parts of context for automatic source code summarization~\cite{su2024distilled}.

\vspace{-0.3cm}
\subsection{Interface}
\label{sec:interface}
We show a screenshot of our study interface in Figure~\ref{fig:interface} with the following notable features:
\begin{itemize}
    \item We built our interface within the Eclipse IDE, using iTrace plugin~\cite{behler2023itracetoolkit} and core~\cite{shafferiTrace}.
    \item Study Instructions are in a markdown file named StudyInstructions.md, which contains an example of a context-aware summary.
    \item For each project, we provide the participant with \textit{project}.txt file, which contains 8 tasks. For each task, we provide the path to the Java file, the name of the method, the line number, and space for the description.
\end{itemize}

\vspace{-0.3cm}
\subsection{iTrace}
\label{sec:iTrace}
We use iTrace suite~\cite{shafferiTrace,iTracetool} for our eye tracking study. The suite includes three tools: 1) \textbf{iTrace core}~\cite{guarnera2018iTrace} that manages the eye tracking, screen recording, keyboard, and mouse recording; 2) \textbf{iTrace Eclipse plugin}~\cite{clark2017iTracevis} that connects to the core and maps the gaze data to specific elements of the IDE and the text inside files; 3) \textbf{iTrace toolkit}~\cite{behler2023itracetoolkit} which is a post-processing tool to isolate gaze patterns.
We use this suite because it can be used in an IDE and can record eye tracking data over the entire code project.

\vspace{-0.3cm}
\subsection{Study Procedure}
We ask each participant to complete five sessions, each with the following study procedure: 
\vspace{-0.1cm}
\begin{itemize}

\item Before the participant arrives, we launch both the Tobii eye tracker manager and iTrace core program. We then set up the session and participant details.

\item Next, the study administrator sets up the Eclipse IDE. First, we connect the iTrace Eclipse plugin to the iTrace core. Second, we set the project explorer to only show the project archived by srcML~\cite{collard2011lightweight} for that session. Then, we open two files: 1) studyInstructions.md and 2) \textit{project}.txt, where \textit{project} is a placeholder for the name of the Java project. See Section ~\ref{sec:javaproj} for details.

\item Next, the study administrator seats the participant. We ask the participant to read the StudyInstructions.md file. These instructions contain an example of the three-sentence summary we ask the participants to write. This step is required for the first session, but the file is always accessible through the session and for future sessions if the participant needs a reference.

\item Next, we ask the participant to open the \textit{project}.txt file that describes the project and contains the location of the 8 Java methods. Each method is a \emph{task} for the purposes of this study.

\item We ask the participant to perform two calibrations before starting the first task as recommended by iTrace developers~\cite{guarnera2018iTrace}. First, Tobii eye tracker manager and then the iTrace core calibration. After calibration, we ask the participant to start a tracking session on the iTrace core window and begin the task inside the Eclipse IDE.

\item To complete a task, we ask the participant to navigate to a file and line number and find the method. Then, the participant is free to explore any number of files in the project. After the completion of one task, we designed the text file to remind the participant to stop tracking and start a new tracking session for the next task. 
This allows each task to have its own eye tracking data and screen recording file. 

\item The study administrator helps re-calibrate with both Tobii eye tracker manager and iTrace halfway through the session, after task 4, and before starting the new tracking session for task 5. We also prompt re-calibration if the participant needs a break or leaves the eye tracking room at any point during the session. 
\end{itemize}

\begin{table*}
\vspace{-0.3cm}
\centering
\caption{Description of Java Projects}
\label{tab:projectstats}
\vspace{-0.35cm}
\begin{tabular}{l|l|l|l|l|l}
\textbf{Name}& Number of Files & Number of Classes & Lines of Code & Age of Project (years) & Domain            \\ 
\hline
Scrimage    & 122             & 111               & 5981          & 11            & Image Editing     \\
MLTK        & 158             & 152               & 18816         & 2             & Machine Learning  \\
OpenAudible & 107             & 80                & 13379         & 6             & Multimedia        \\
MALLET      & 619             & 767               & 77407         & 16            & NLP               \\
FreeCol     & 804             & 889               & 128049        & 20            & Entertainment    
\end{tabular}
\end{table*}

\vspace{-0.3cm}
\subsection{Java Projects}

\label{sec:javaproj}
We scraped Github for repositories with at least 90\% of the project code base containing Java source code. We also excluded projects that depend heavily on other projects or codebases. Finally we picked the following five projects: 
\begin{itemize}
    \item \textbf{Scrimage} is an immutable, functional, and performant Java library for manipulation of images. The aim of this library is to provide a simple and concise way to do common image operations, such as resizing, filter, and converting between formats.
    \item \textbf{MLTK} is a collection of various supervised machine learning algorithms, which is designed for directly training models and further development.
    \item \textbf{OpenAudible} is a cross-platform desktop application for downloading and managing Audible audiobooks with both a GUI interface and an organizational library.
    \item \textbf{MALLET }is a Java-based package for statistical natural language processing, document classification, clustering, topic modeling, information extraction, and other machine learning applications to text.
    \item \textbf{FreeCol} is a turn-based strategy game based on the old game Colonization, and similar to Civilization. The objective of the game is to create an independent nation.
\end{itemize}
More details about the projects can be found in Table \ref{tab:projectstats}. Notably, the projects cover a wide range of project domains, an 18 year age range, and between 5-128k lines of code. For each project above we chose 8 specific methods. To ensure that the methods can be described using relevant context, we avoided trivial methods such as getters, setters, or similar self-evident functions, while prioritizing methods that are called at least twice (for multiple direct contextual sources) or perform some functionality in the project discernible in the context of the project. The order of methods 2 through 8 is randomized. We do this to answer RQ1 by minimizing the effect of any one method on the observations, i.e., a particularly tough or easy method would not skew the results as it would appear at a different position for each participant. Method 1 is fixed for each session to start each participant off with the same baseline.

\vspace{-0.25cm}
\subsection{Eye tracker}
\label{sec:hardware}
The eye tracker used for this study was the Tobii Pro Fusion at 120Hz mounted at the bottom of a Spectre 24-inch 1920x1080 resolution monitor at 60Hz screen refresh rate. The eye tracker has an accuracy of 0.03$^{\circ}$ and precision of 0.04$^{\circ}$ in optimal conditions~\cite{tobii}. Note, although all sessions were conducted in the same lab with the same equipment, we could not ensure optimal conditions were met for all participants due to factors such as ambient light, natural light and corrective lenses.

\vspace{-0.25cm}
\subsection{Software Versions}
We list the software versions we used for our study below to promote future replication of the study and results: 
Windows 10 pro 10.0.19045; iTrace core v0.2.0; iTrace Eclipse Plugin v0.2.0; iTrace Toolkit 0.2.2; Eclipse IDE for Java V2023-06 (4.28.0); OpenJDK 18

\vspace{-0.25cm}
\subsection{Participants}
\label{sec:participants}
We recruited 10 programmers with Java experience to participate in our study. We recruited participants by email list, and we collected informed consent which clearly indicated they could leave at any time, and were compensated 60 USD per session. Each participant was required to have at least 1 year of previous Java development experience. The participants had an average of 2.8 years of Java development experience and 5 years of general programming experience. Each participant was compensated at a flat rate of USD 60 per session, with an average duration of around 1 hour per session. 4 participants identified as female, and 6 as male. 6 participants identified as non-native but fluent English speakers, and 4  as native English speakers. We designed the study to be in-depth to answer RQ1, where each participant processes multiple methods and projects, so we did not recruit a high number of participants. In this study we focus on general findings instead of individual differences.

\section{Methodology}
In this section, we discuss the methodology for analysis of data after the eye tracking study.

\vspace{-0.25cm}
\subsection{Gaze Pattern Metrics}
Following the good practices established by similar studies in the past~\cite{buschersurfing,booth2013function,rodeghero2015empirical}, we calculate and report the \textbf{statistical summaries} of the following gaze pattern metrics to answer the three RQs:

\begin{description}
\item[Fixation:] A fixation is marked by a steady gaze of the human eye at a particular location in the stimulus, for the duration of at least 100-200ms, required for the visual system to process information. It is the most common measure of human visual attention in eye tracking literature~\cite{buschersurfing}.  For this study, we only compute statistical summaries of fixation count and duration.

\item[Regression Rate:] A regression occurs when a fixation occurs against the current direction of reading, such that a previously read word may be fixated upon again, marking an event of re-reading~\cite{booth2013function}. Higher regressive rate may indicate higher difficulty in understanding~\cite{son2022attention}. We calculate regression rate for each method task as the number of regressive fixations normalized by the number of total fixations during that task.

\item[Lines of code:] We compute and report statistical summaries of counts, density, time spent, and transition between lines of code~\cite{rodeghero2015empirical} seen by a participant as a measure of programmer attention.

\item[Methods visited:] We compute and report statistical summaries of counts, time spent, and manually labeled relationships between methods in the project as a measure of programmer attention. The idea is that the methods that linked to the highest gaze activity, receive more attention from the programmer.
\end{description}

\vspace{-0.35cm}
\subsection{Qualitative Annotation}
\label{sec:grading}
Two of the authors annotated the quality of summaries generated by the participants, to assess the effect of changing gaze patterns on performance, based on four qualities recommended in related works~\cite{treude2020beyond}:
\begin{itemize}
    \item~\textbf{Accuracy} On a scale of 1-5, how accurate is the summary? (deduct a point for each instance of inaccuracy)
    \item~\textbf{Completeness} On a scale of 1-5, how complete is the summary? (deduct a point for each instance of missing important information)
    \item~\textbf{Conciseness} On a scale of 1-5, how concise is the summary? (deduct a point for each instance of unnecessary information)
    \item~\textbf{Clarity} On a scale of 1-5, how clear or readable in the summary? (deduct a point for each instance of lack of clarity or grammatical incorrectness)
\end{itemize}

 When assigning annotations or ``codes'' in a qualitative manner, there are two popular approaches to reduce human biases. The first approach is an~\emph{agreement-based approach}~\cite{rastkar2014automatic}, which relies on a high number of annotators (or graders for the purpose of this study) to reach a Kappa agreement threshold. However, that approach does not account for expertise, i.e., a large number of novices may override an expert. For our study, every grader would have to be 1) expert in Java documentation, and 2) familiar with the entirety of the project, as we expect our participants to explore the project while writing summaries. The second approach is a~\emph{\textbf{discussion-based approach}}, which relies on an iterative process of discussion for each disagreement. We chose this approach as recommended by related work for tasks with a low tolerance for imperfect grading~\cite{rieser2011reinforcement,wood2018detecting}. Both graders independently graded all summaries, followed by a discussion phase on each disagreement, until both graders agreed on one score. To avoid agreement bias, the graders did not know the participant number or the order in which the summaries were shown to the participants.

\vspace{-0.3cm}
\subsection{Context Categorization}
We divide all areas of context in any of the Java projects into five categories as they relate to the target method:
\begin{description}
    \item[Class or Instance methods:] are the areas of context that fall within a method in the same class as the target method.
    \item[Class declarations:] are the areas of context that fall within the same class as a target method but not inside a method, such as object and variable declarations.
    \item[Call graph:] are the areas of context that fall within a method in the call graph of the target method, i.e., it is a caller or callee of the target method.
    \item[File context:] are the areas of context that fall within the same Java file as the target method but are not in the same class as the target method or in the call graph. These may be import statements, global variable declarations or other private classes defined in the same Java file.
    \item[Project context:] are all other areas of context in the project that are not included in any of the above categories. 
\end{description}

We use srcML~\cite{collard2011lightweight} to automatically identify the type of source code element -- such as conditional or method call -- in each project. iTrace~\cite{shafferiTrace} calculates the file, column, and row data from the eye tracking data, which we use to identify the source code element the participant was looking at. From this we use srcML to identify the area of context in which the source code element is at each point.

\section{Results}
\label{sec:results}

We report the results for our three RQs in this section.  For perspective, key statistics about the dataset about the dataset include participants, gazes, and fixations per task:

\begin{table}[h]
\centering

\begin{tabular}{l|l|l|l}
Participants & 10 & Fixations (Mean) & 263.5 \\
Hours/Participant & 6 & Fixations (Min) & 12 \\
Total Fixations & 95924 & Fixations (Max) & 1726 \\
Total Gazes & 143570 & & \\
\end{tabular}
\end{table}
\vspace{-0.2cm}

\begin{table*}[t!]
\centering
\normalsize
\caption{Mean and Median values for metrics when grouped by the position in which the method was seen.}
\label{tab:rq1}
\vspace{-0.3cm}
\begin{tabular}{l|rl|rl|ll|rl|rl}
Method & \multicolumn{2}{c|}{Fixation Count}                    & \multicolumn{2}{c|}{Avg. Fixation Duration}                 & \multicolumn{2}{c|}{Regression Rate}                  & \multicolumn{2}{c|}{Lines Visited}                     & \multicolumn{2}{c}{Methods Visited}                   \\ \cline{2-11} 
& \multicolumn{1}{c}{Mean} & \multicolumn{1}{c|}{Median} & \multicolumn{1}{c}{Mean} & Median & \multicolumn{1}{c}{Mean} & \multicolumn{1}{c|}{Median} & \multicolumn{1}{c}{Mean} & \multicolumn{1}{c|}{Median} & \multicolumn{1}{c}{Mean} & \multicolumn{1}{c}{Median} \\ \cline{1-11}
1      & 331.61                   & 259                        & 444.39                   & 289                        & 56.65                    & 56.93                      & 50.03                    & 38.5                       & 9.04                     & 5                          \\
2      & 268.54                   & 249.5                      & 480.37                   & 300                        & 53.78                    & 53.48                      & 46.92                    & 45                         & 9.71                     & 6.5                        \\
3      & 302.89                   & 221                        & 441.05                   & 281                        & 55.07                    & 56.94                      & 45.45                    & 38                         & 9.11                     & 7                          \\
4      & 290.46                   & 250.5                      & 484.24                   & 302                        & 56.21                    & 59.21                      & 40.58                    & 34                         & 9.35                     & 8                          \\
5      & 238.77                   & 216                        & 491.90                   & 310.5                      & 52.89                    & 52.39                      & 40.58                    & 36.5                       & 10.08                    & 7                          \\
6      & 202.81                   & 156                        & 458.15                   & 283                        & 51.1                     & 49.67                      & 36.03                    & 32                         & 8.48                     & 6                          \\
7      & 222.44                   & 193                        & 478.40                   & 301.5                      & 51.52                    & 54.78                      & 35.08                    & 31                         & 6.96                     & 7                          \\
8      & 214.69                   & 176                        & 491.89                   & 308                        & 49.85                    & 51.97                      & 40.81                    & 30.5                       & 8.19                     & 6                       
\end{tabular}
\vspace{-0.2cm}
\end{table*}

\subsection{RQ1: Effect of the progress through tasks}
\label{sec:rq1}

We divide our discussion of the answer to RQ1 into the following topics: Overall Information, Key Observations, Statistical Tests, and Interpretation. 
\vspace{.25cm}

\begin{table}[b!]
\vspace{-0.3cm}
\normalsize
\centering
\caption{Results of the Mann-Whitney U statistical tests on the distributions presented in Table ~\ref{tab:rq1}}
\vspace{-0.3cm}
\label{tab:rq1stats}
\begin{tabular}{l|l|l|c}
{Gaze Metrics}        & \multicolumn{3}{c}{Mann-Whitney U} \\
           & U1      & U2     & p-value       \\ \cline{1-4}
Fixation Count    & 1892.5  & 1075.5 & \textbf{0.01} \\
Avg. fixation duration & 1263    & 1705   & 0.18          \\
Regression Rate   & 1193.5  & 1774.5 & \textbf{0.02} \\
Lines Visited     & 1773.5  & 1194.5 & 0.08          \\
Methods Visited   & 1501    & 1467   & 0.92         
\end{tabular}
\end{table}

\begin{table}[b!]
\caption{Average quality ratings for summaries when grouped by the position of the methods.}
\label{tab:rq2method}
\begin{tabular}{c|c|c|c|c|c}
Method & Accurate & Complete & Concise & Clear & Overall \\ \cline{1-6}
1                    & 4.32                         & 3.94                             & 4.42                            & 4.74                        & 4.36                        \\
2                    & 4.36                         & 4.10                             & 4.40                            & 4.40                        & 4.32                        \\
3                    & 4.46                         & 4.12                             & 4.22                            & 4.64                        & 4.36                        \\
4                    & 4.44                         & 4.22                             & 4.28                            & 4.58                        & 4.38                        \\
5                    & 4.56                         & 4.26                             & 4.28                            & 4.54                        & 4.41                        \\
6                    & 4.45                         & 3.98                             & 4.29                            & 4.63                        & 4.34                        \\
7                    & 4.50                         & 4.10                             & 4.13                            & 4.65                        & 4.34                        \\
8                    & 4.32                         & 3.98                             & 4.09                            & 4.68                        & 4.27                        \\ \cline{1-6}
\textbf{All}     & 4.43			       &4.09				    &4.26				 & 4.61			 & 4.35                        \\
\end{tabular}
\end{table}

\begin{table*}[t!]
\setlength{\tabcolsep}{4.5pt}
\centering
\normalsize
\caption{Gaze metrics for summaries grouped by quality scores: low(\textless=3) and high(=5) for completeness and conciseness. Here U1, U2, and p-value are results of the Mann-Whitney U statistical test for each group.}
\label{tab:rq2}
\vspace{-0.2cm}
\begin{tabular}{l|rr|rr|rr|rr|rr}
Group                     & \multicolumn{2}{c|}{Fixation Count}                         & \multicolumn{2}{c|}{Regression Rate}                        & \multicolumn{2}{c|}{Avg. fixation duration}                      & \multicolumn{2}{c|}{Lines Visited}                          & \multicolumn{2}{c}{Methods Visited}                        \\ \cline{2-11}
                          & \multicolumn{1}{l}{Complete} & \multicolumn{1}{l|}{Concise} & \multicolumn{1}{l}{Complete} & \multicolumn{1}{l|}{Concise} & \multicolumn{1}{l}{Complete} & \multicolumn{1}{l|}{\,\quad  Concise} & \multicolumn{1}{l}{Complete} & \multicolumn{1}{l|}{Concise} & \multicolumn{1}{l}{Complete} & \multicolumn{1}{l}{Concise} \\ \cline{1-11}
Low(\textless{}=3) & 181.72                       & 350.38                      & 48.38                        & 55.78                       & 562.08                       & 475.47                      & 29.88                        & 53.62                       & 5.60                         & 11.42                       \\
High(=5)           & 264.23                       & 228.87                      & 54.27                        & 52.55                       & 482.01                       & 516.73                      & 43.25                        & 38.49                       & 9.29                         & 8.14                        \\ \cline{1-11}
U1            & 1,856.00                     & 2,765.00                    & 2,239.00                     & 2,353.00                    & 2,764.00                     & 1,825.00                    & 1,740.50                     & 2,743.50                    & 1,796.00                     & 2,684.00                    \\
U2            & 3,734.00                     & 1,369.00                    & 3,351.00                     & 1,781.00                    & 2,826.00                     & 2,309.00                    & 3,849.50                     & 1,390.50                    & 3,794.00                     & 1,450.00                    \\
p-value                   & 0.001                        & 0.006                       & 0.051                        & 0.259                       & 0.915                        & 0.340                       & 0.000                        & 0.008                       & 0.000                        & 0.015                      
\end{tabular}

\end{table*}

\textbf{Overall Information}
We found that participants need less information to complete their context-aware summarization task as they progress through the tasks. To analyze the change in gaze patterns during this process, we report the change in average fixation duration, fixation count, regression rate, lines visited, and methods visited by the participants. We make two interesting observations.
We found that participants need less information to complete their context-aware summarization task as they progress through the tasks. To analyze the change in gaze patterns during this process, we report the change in average fixation duration, fixation count, regression rate, lines visited, and methods visited by the participants.

\textbf{Key Observations}
First, we observe a pattern of decreasing fixation counts. In Table~\ref{tab:rq1}, we report the statistical summaries of the ~\emph{count} of fixations when grouped by position of the method, which denotes the position in which the participants saw and wrote summaries for the method. We observe a 35\% decrease in the mean number of fixations the participants made on the last method when compared to the first method processed. However, with respect to average fixation duration, we did not observe a significant change. This may be because average time spent on a fixation may be affected by the participants' reading speed.

Second, we also observe a clear trend of decreasing regression rate. The mean of normalized regression rates for each task decreased by roughly 13\% when compared by method position. Regressions often occur when information is particularly hard to understand or needs context from the surrounding words before it can be understood~\cite{liu2020mind}. This indicates that as the participants progressed through the session they did not encounter word tokens that were hard to comprehend or needed additional information to understand as frequently as when they started.

\textbf{Statistical Tests}
We performed the Mann-Whitney U~\cite{nachar2008mann} test, a non-parametric, non-paired statistical test on the distribution of the values, and report the results in Table~\ref{tab:rq1stats}. We chose this test because our data does not meet the assumptions of a parametric test, i.e., it is subjective and is not known to follow Gaussian or Normal distributions. We compared the combined distribution of the first two methods against the combined distribution of the last two methods seen by the participants. We observed that for fixation counts the difference between the two distributions were statistically significant ($p<0.01$). We also observed statistically significant difference between the two distributions for regression rate ($p=0.02$)

However, with respect to number of lines visited and number of methods visited, we observe a p-value $>0.05$. Therefore, we cannot reject the null hypothesis for these metrics. There is statistical likelihood that patterns such as decreasing mean number of lines visited may not be significant. These observations are interesting because they indicate that the participants continued to cover a large amount of the context to write these descriptions as they processed more methods.

\textbf{Interpretation}
Our interpretation of these observations is that the participants retain more information about the project from previous task, and thus needed less information to write summaries as they processed more tasks. Different studies have shown how programmers skim source code whenever possible, reading the minimum amount of information they need~\cite{starke2009searching}. As programmers become more familiar with the task, they are more likely to intuitively know what information they need, and therefore read fewer details. Our observations support those findings as we do not see a significant decrease in lines and methods visited by the participants but fixations and regressions decrease. Therefore, participants still covered similar amount of code but seemed to get better at skimming, resulting in lower frequency of fixations and regression rate.

An alternative explanation might be fatigue or boredom, the effects of these are marked by decreased thoroughness or decreased quality of summaries written. Our observations contradict the former as we did not observe statistically significant effect on code context coverage in terms of the number of lines and methods visited. We study the possibility of the latter in the next section about RQ2.


\vspace{-0.2cm}
\subsection{RQ2: Quality of Summaries}
\label{sec:rq2}

To analyze the quality of summaries written by the participants, we had two graders manually grade each summary for accuracy, completeness, conciseness, and clarity (see Section~\ref{sec:grading} for details). We make a the following observations.

\begin{figure*}[t!]
	\centering
	\begin{tabular}{cc}
        \multicolumn{2}{c}{\includegraphics[width=0.75\textwidth] {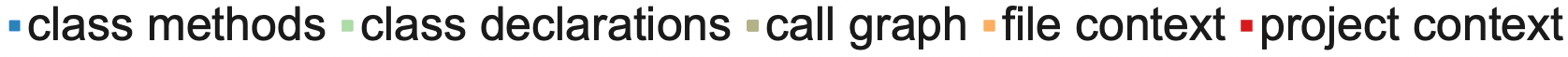}\vspace{-0.15cm}}\\
		\hspace*{-0.5cm}\begin{tabular}{l}\includegraphics[width=0.5\textwidth]{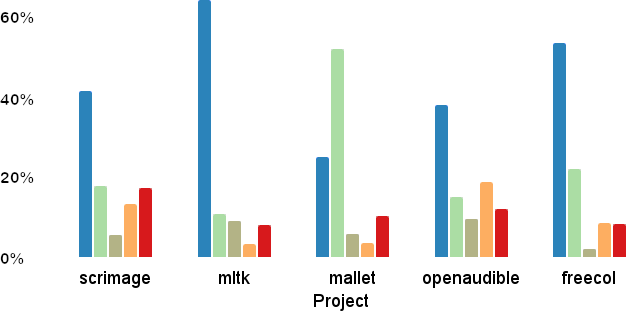}\end{tabular}  
        &\hspace*{-0.4cm}\begin{tabular}{l}\includegraphics[width=0.5\textwidth]{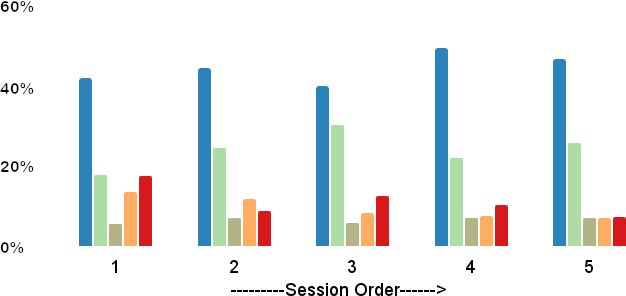}\end{tabular}\vspace{-2mm} \\
       
		{\small \texttt{(a)}}  & {\small \texttt{(b)}} \\
		\hspace*{-0.45cm}\begin{tabular}{l}\includegraphics[width=0.5\textwidth]{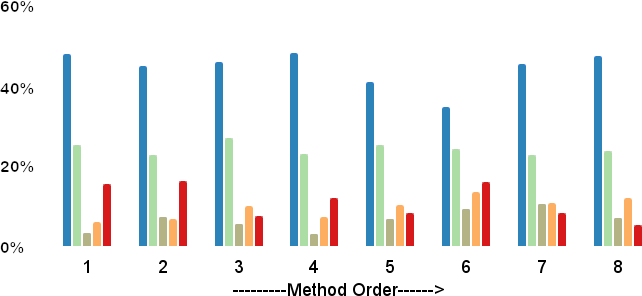}\end{tabular}  
        &\hspace*{-0.4cm}\begin{tabular}{l}\includegraphics[width=0.5\textwidth]{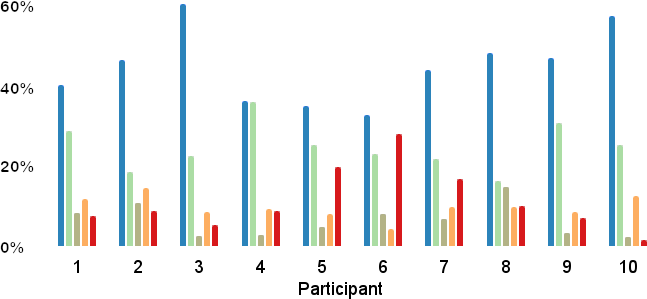}\end{tabular}\vspace{-2mm} \\
        
		{\small \texttt{(c)}}  &{\small \texttt{(d)}}  \\
	\end{tabular}
	\vspace{-0.5cm}
	\caption{Graphs illustrating the distribution of programmer attention for varying groups defined on the X-axis as (a) project names, (b) the order in which the session occurred, (c) the order in which the method was seen, and (d) the participant ID. The legend on top is common for the bar colors on all graphs. The Y-axis in each graph shows mean value for each type of context. The values were normalized by the total time spent fixating on context outside the target method for each method-summary pair, prior to computing mean for each category.}
	\label{fig:graphs}
	\vspace{-0.45cm}
\end{figure*}

First, we observe no significant pattern associated with the position in which the summary was written. In Table~\ref{tab:rq2method}, we report the mean scores for the four quality metrics. We also report an over score, an average of those 4 metrics calculated individually for each summary. We also performed the Mann-Whitney U test between the distributions of the first two methods and the last two methods in a way consistent with RQ1, with an observed $p\ge0.9$. This observation supports our hypothesis about RQ1 that a decrease in fixation counts and regression rate is likely not due to exhaustion, as neither code context coverage nor quality of summaries was affected by the position in which the methods were seen. 

Next, we found that with $mean=4.35$ and $median=4$ on a scale of 1-5, the summaries were graded to be of high quality.
We observed a higher variance in scores for two of the qualities, completeness and conciseness, compared to accuracy and clarity.
Therefore, we analyze the eye gaze patterns between two groups, those that received a low score ($\textless{}3$) for completeness ($n=44$) and conciseness ($n=33$), against those that received a high score ($=5$) for completeness ($n=147$) and conciseness ($n=170$). Here $n$ denotes the number of summary samples in each subset. We do not compare groups based on accuracy and clarity because the low score group size is too small ($n\textless{}5$). 
\begin{figure}[b!]
    \vspace{-0.4cm}
	\centering
	\includegraphics[width=\linewidth]{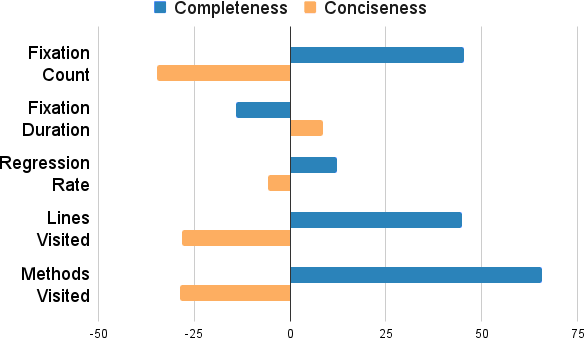}
	\caption{Bar chart showing the delta values for gaze metrics computed between summaries rated low(\textless{}=3) and highly(5) in terms of completeness and conciseness.} 
	\label{fig:qualitydelta}
\end{figure}

We observe that the completeness of summaries is correlated to the code context coverage by the programmer. In Table~\ref{tab:rq2}, we report the mean values of each of the five gaze metrics for each subset. We also report the U1, U2, and p-value for Mann-Whitney U 
statistical test. We observe that the summaries of high completeness correspond to a \emph{significantly} ($p < 0.001$) higher number of fixations, lines visited, and methods visited. Figure~\ref{fig:qualitydelta} shows this delta clearly through the blue bars. This suggests that the participants that covered a smaller amount of the code context, wrote summaries that were missing important information. 
Although we observe an increase in mean regression rate as well, the p-value for regression rate $p=0.051$, which is not statistically significant by conventional standards, but approaches significance. 
We do not observe a significant difference in average fixation duration. This observation supports the observation in RQ1, and further suggests that average fixation duration may be dependent on each participant's reading and comprehension speed. 

On the other hand, the conciseness of summaries is inversely correlated to the code context coverage by the programmer. From Table~\ref{tab:rq2}, we observe that the summaries of high conciseness correspond to a~\emph{significantly} ($p<0.015$) lower number of fixations, lines visited, and methods visited. Figure~\ref{fig:qualitydelta} shows this delta more clearly. This suggests that participants included unnecessary information at a higher rate when they covered an exceptionally larger part of the code context. As shown in Figure~\ref{fig:qualitydelta}, we did not observe a significant difference in average fixation duration or regression rate between the two groups.

Our interpretation of these observations is that there are diminishing returns in terms of increasing code context coverage and improvement in the quality of summaries. We found that gaze metrics for the summaries that achieved a perfect score for all four qualities ($Overall=5$), tend to fall between the thresholds for high completeness and conciseness. For example, in terms of lines visited, the summaries with a perfect score have a mean score of 41 lines, which is between the scores of 43.25 and 38.49 reported in Table~\ref{tab:rq2}. 

In contrast, summaries marked for low completeness have a lower mean score of 29.88, while summaries marked for low conciseness have a high mean score of 53.62, indicating that reading more context does not always result in a better summary. Rodeghero~\emph{et al.}~\cite{rodeghero2014improving} and Abid~\emph{et al.}~\cite{abid2019developer} suggested that function signatures or call terms (such as method name and parameter) are key components of code in the summarization process, indicating some lines of code that contain more information than others, aligning with our finding that code context is useful for complete summaries, but concision decreases with lines read. 

So far, we have looked at how gaze patterns of participants change in relation to the task, the quality of summaries generated, and the amount of context analyzed. In the next subsection, we categorize the type of context on which the participants fixated.

\vspace{-0.25cm}
\subsection{RQ3: Attention to Context }
\label{sec:rq3}

We found that of the types of context, participants spent a vast majority of their time looking at contextual methods and variable declarations in the same class as the target method. Surprisingly, we found that participants spent a comparatively small amount of time looking at methods in the call graph, even though we designed the second and third sentences in the summary to contain information from the callees and callers respectively. In Figure~\ref{fig:graphs} we report four graphs, each showing averages for total time spent fixating on tokens inside each type of context, normalized by total time spent looking at context for the task. 

In Figure~\ref{fig:graphs} \textbf{(a)}, we report the attention grouped by the project type. We found see similar trends for four of the five project, namely Scrimage, Openaudible, MLTK, and FreeCol. For these projects, participants spent an average of 40-65\% of their time looking at methods in the same class as the target method. 
Surprisingly, for the Mallet project, on an average participants spent more time looking at the variable, declarations, and other data inside the class than methods and calls. One possible explanation for this outlier we investigated was that Mallet relies more on class variables than the other projects. However, we did not find significant differences between the number of class declarations and variables for Mallet, compared to other projects.

In Figure~\ref{fig:graphs} \textbf{(b)}, we analyze the difference in attention grouped by session, i.e., the order in which the project was processed by a participant. We found that on average as participants processed more sessions, they relied more on information inside the class, both methods and class declarations. This is expected because as participants repeat the task, they 
form a consistent pattern of navigating the context. This supports the earlier observation that overall, the participants found the class of the target method to be most important part of the context.

In Figure~\ref{fig:graphs} \textbf{(c)}, we analyze the difference in attention grouped by method, i.e., the order in which the method was processed by a participant in any given session. We observe no significant trend based on the order in which the method was processed. This supports our observation from RQ1 and RQ2 that although the participants make fewer fixations as they process more methods, they still get the same general code context coverage in the context.

In Figure~\ref{fig:graphs} \textbf{(d)}, we analyze the difference in attention of each participant. We found that most of the participants relied on class and instance methods and the class declaration for contextual information. We found that only participant 5 and 6 are outliers, and spent a significant amount of time fixating on context in other files, and these are areas that do not fall under the methods in the call hierarchy. This suggests that observations from \textbf{(a)}, \textbf{(b)}, and \textbf{(c)} are consistent with a majority of the participants and not skewed heavily by an extreme minority.

\newpage

Overall, our interpretation of these results is that there is a general agreement of trends between the participants, which is unaffected by method or session order. We observe a few spikes in class declarations, which may be skewed by the Mallet project. These observations make a case that the most important part of context for a summarization task like this is all the information inside the class.

\vspace{-0.2cm}
\section{Example}

\begin{figure}[b!]
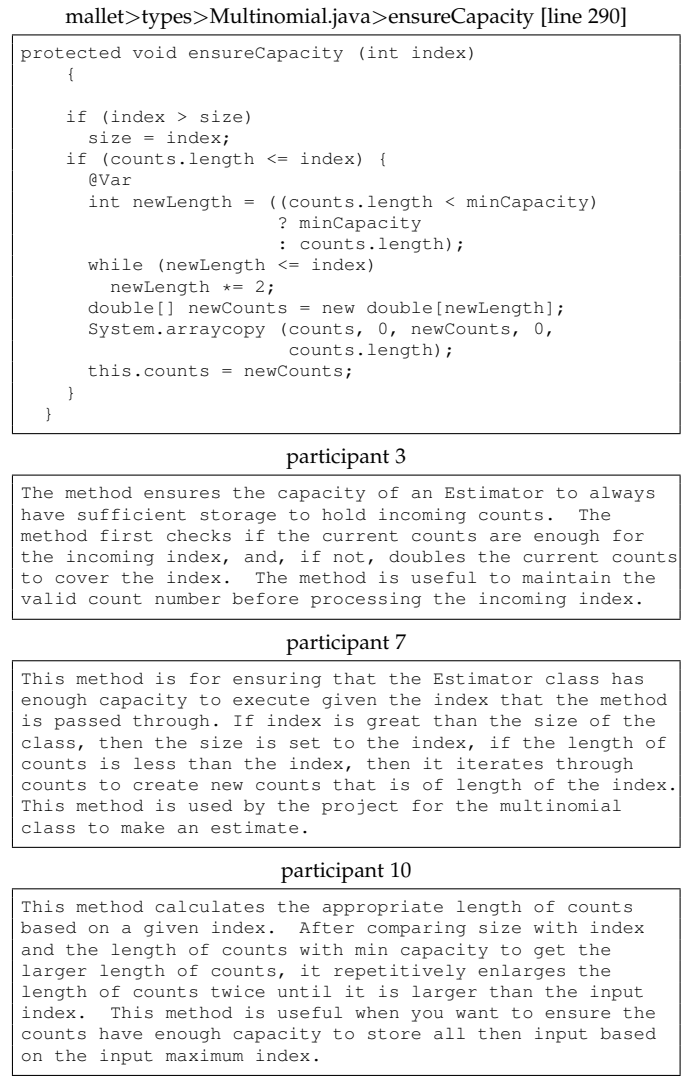

\centering
\footnotesize
\vspace{-0.5cm}
mallet.types.Multinomial.java.ensureCapacity [line 290]
\begin{Verbatim}[fontsize=\scriptsize,tabsize=2,frame=single]
protected void ensureCapacity (int index)
    {
			
		if (index > size)
			size = index;
		if (counts.length <= index) {
			@Var
			int newLength = ((counts.length < minCapacity)
							         ? minCapacity
							         : counts.length);
			while (newLength <= index)
				newLength *= 2;
			double[] newCounts = new double[newLength];
			System.arraycopy (counts, 0, newCounts, 0, 
                        counts.length);
			this.counts = newCounts;
		}
	}
\end{Verbatim}
\vspace{-1mm}
Participant 3: \hfill Accuracy: 5, Completeness: 5, Conciseness: 5, Clarity: 5
\begin{Verbatim}[fontsize=\scriptsize,tabsize=2,frame=single]
The method ensures the capacity of an Estimator to always 
have sufficient storage to hold incoming counts.  The 
method first checks if the current counts are enough for
the incoming index, and, if not, doubles the current counts 
to cover the index.  The method is useful to maintain the 
valid count number before processing the incoming index.
\end{Verbatim}
\vspace{-1mm}

Participant 4: \hfill Accuracy: 5, Completeness: 4, Conciseness: 2, Clarity: 3
\begin{Verbatim}[fontsize=\scriptsize,tabsize=2,frame=single]
This method ensures that a double array counts has the
correct capacity for further use in estimation operations. 
The method receives an index and checks if the double array 
needs to be resized because the index exceeds the current 
size. If it does exceed, a new array is created with the
new size and an array copy operation takes place before 
setting the original counts array to the temporary one.  
This utility function is useful for determining when an
array needs to be resized with further capacity. By 
checking capacity before resizing, the program can ensure 
memory is not being wasted
\end{Verbatim}
\vspace{-1mm}

Participant 10: \hfill Accuracy: 5, Completeness: 5, Conciseness: 4, Clarity: 5
\begin{Verbatim}[fontsize=\scriptsize,tabsize=2,frame=single]
This method calculates the appropriate length of counts 
based on a given index.  After comparing size with index 
and the length of counts with min capacity to get the 
larger length of counts, it repetitively enlarges the 
length of counts twice until it is larger than the input
index.  This method is useful when you want to ensure the
counts have enough capacity to store all then input based 
on the input maximum index.
\end{Verbatim}
\vspace{-0.5cm}
	\caption{Example with source code of a method in project mallet, accompanied by summaries written by participant 3, 4, and 10. }
	\label{fig:example}
\end{figure}

In Figure~\ref{fig:example} we show an example containing target method and three corresponding summaries written by participant 3, 4, and 10. These may explain some of our observations from Section~\ref{sec:results}. Notice, the summaries written by Participant 3 and 10 are similar, but the summary written by 10 is slightly clearer. On the other hand, the summaries written by Participant 4 for the same method are lengthy and hard to comprehend and received one of the lowest overall scores by our graders (Overall = 3.5).

We found that the key to writing the last sentence for both participant 3 and participant 10's summaries is to track the class variable~\emph{minCapacity}. Instances like this might explain why mallet shows a higher attention on class declarations than methods as observed in Section~\ref{sec:rq3}. 

For this task, 90\% of the participants first looked at the ternary operator and~\emph{arraycopy} elements, before looking through the call graph and class methods. We provide a representative example from participant 3: The participant started with~\emph{minCapacity}, before following the call graph to \textit{incremen}, which is also a class method. After tracking through several overloaded methods, they looked back at the lower half of~\emph{minCapacity} before looking back to the instructions, where they write their summary. After looking back and forth from the target method and the call graph twice more, they move to the next task.

We also notice that participant 4's summary is an example of summaries where excessive code coverage leads to verbose and unclear summary. Note this is just one example we found that fits the observations from Section~\ref{sec:rq2}, we do not claim that this example is representative of a trend or make any recommendations based on this example. 

\vspace{-0.2cm}
\section{Threats to Validity}
Our study has several threats to validity. One threat lies with the selection of projects. We selected open-source Github projects that neither we nor the participants had any contribution in developing. 
Some of our results might be specific to these projects and any errors or shortcomings these projects may have. We tried to mitigate this threat by choosing projects on varying topics, with several contributors and varying sizes.  

Our selection of methods is a threat to validity because, like all studies of this nature, it is not feasible to evaluate all methods and our selection of methods ultimately relies on our judgment (see Section \ref{sec:javaproj}). A different set of methods could lead to different results. The methods chosen could be too simple and not rely on context to understand them. To mitigate this threat, we chose methods that were called multiple times or had roles evident from the context. We also asked the participants in their instructions to explain the method in context of the whole program to encourage them to explore enough of the context to be accurate. 

Another threat lies with our pool of participants. Our participants are graduate students, and their experiences may not translate directly to those of active industry professionals. To mitigate this threat, we screened participants with previous (and recent) Java development experience. However, industry professionals may have specific mental models for their industry and may utilize visual attention strategies that our participants do not represent. Additionally, the participants in our study had a range of experience between two and five years, meaning that it is possible that our results would not generalize to people with experience levels outside this range. Another threat to validity is our sample of participants. Even though we had 60 hours of total programmer time, these hours were from ten programmers.  A different set of programmers could in theory lead to a different set of conclusions. Another threat with our pool of participants is that half of them wear corrective glasses. In specific circumstances, corrective glasses can lead to false fixation readings. To try to mitigate this threat, we recommended that participants complete the task without glasses or use contact lenses if they are able.

Another threat lies with administrators. Our studies were administered by two administrators across several months. There may be differences in the administration styles and the way the task was explained to each participant, which could result in variation between data points. To somewhat mitigate this threat, we created a common script and provided the instructions in text file to each participant. 

Another threat lies with the study conditions. The room in which we conducted the study has a window. As the study was conducted over several months, the lighting conditions may have changed between sessions. Newer generation eye tracker claim to work in all lighting conditions but can cause inconsistencies in rare situations. These external threats exist for all eye tracking studies. 

Another threat lies with the extraction of context categories. The accuracy of these categorizations is limited by the accuracy of both srcML~\cite{collard2011lightweight} and iTrace-toolkit~\cite{behler2023itracetoolkit}. Additionally, the categorization of call graph is limited to a static call graph generated by Call-Hierarchy Plugin for Eclipse for Java IDE 2023. We only report average fixation duration and counts, normalized by the sum of fixation duration and counts for a session, to mitigate some of these threats. We also take measures to avoid false positives within any category of context, such that, if a fixation cannot be confirmed to be within the strict boundaries of any of the other categories, we assign it the ``project context'' category.

\section{Conclusion \& Future Work}

In conclusion, we designed an eye tracking study to analyze how programmers read source code during context-aware source code summarization. We first studied how the gaze patterns change as participants progress through tasks. We observed a statistically significant decrease in fixation counts and regression rates as participants summarize more methods in a session. This observation meant that the participant retained and internalized information about the project code base from earlier tasks. Another possibility was that the participant was bored or exhausted. To test the latter, we rated and compared the quality of summaries and found no significant change in the quality of summaries, as one would expect from exhaustion or boredom. 

We also found that regardless of progress, the amount of context a participant looked at in terms of lines visited, methods visited, and fixations affects the quality of summaries. This effect can be best described as similar to a bell curve, such that the quality of a summary increases with coverage of context until a peak, after which the quality declines as the summaries become less concise. The outcome of this effect starts to answer some open questions about what parts of the context people need to understand the rationale behind code, and indicates that finding specific areas of context may be more important than the quantity of context. 

Next, we categorize and analyze the context on that the participants looked at for the longest fixation duration. We found that participants spent the longest time looking at class and instance methods, followed by class declarations and call graph. We observed this trend to be unaffected by project name, task order, session order, or participant.

\vspace{5mm}

This study also contributes to a body of work answering open questions about how quickly people may learn a new codebase, by quantifying that people need to read up to 35\% fewer words after summarizing methods in each session. It also adds to an ongoing discussion in the literature about the need for context to understand code by providing information about not only how much context people need, but also how often they read it.

Based on our study, we propose three main lines of inquiry for future work:

\begin{enumerate}
    \item Based on our findings in RQ1, we propose that more in-depth studies such as ours may be needed to study the effect of retention of project-level information for SE tasks such as code debugging and testing. 
    \item Based on our findings in RQ2 and RQ3, we propose that automated code summarization models may be improved by using specific areas of the project as context. Based on our findings, we recommend future studies prioritize class and instance methods first, class declaration second, and call graph third.  Additionally, analysis of scan paths could be used to discover which reading order results in the best summaries~\cite{Sharafi2020PracticalGuide}.
    \item We hope our dataset of context-aware summaries aids future studies in automatic code summarization. We propose using the results of this study as a guide for generating code summaries that describe rationale, such as showing what code context should be included as inputs for AI-based technologies that generate code summaries or helping decide how large training data window sizes should be. Additionally, we hope this dataset will be useful as a starting point for research in other areas expanding into context-aware AI-driven approaches, such as bug localization. 
\end{enumerate}
\textbf{Reproducibility} To ensure reproducibility of the results, we release all datasets and source code, along with an online reproducibility guide with instructions outlining how we produced our results via an online appendix on Zenodo:

\url{https://doi.org/10.5281/zenodo.14873017}

\vspace{-0.3cm}
\section{Acknowledgments}
This work is supported in part by the NSF grants CCF-2100035, CCF-2211428 and CCF-2211429. 
Any opinions, findings, and conclusions expressed herein are the authors’ and do not necessarily reflect those of the sponsors. We also sincerely thank participants of our qualitative study.

\bibliographystyle{IEEEtran}
\bibliography{IEEEabrv,main}

\vspace{-2cm}

\begin{IEEEbiography}[{\includegraphics[width=1in,height=1.25in,clip,keepaspectratio]{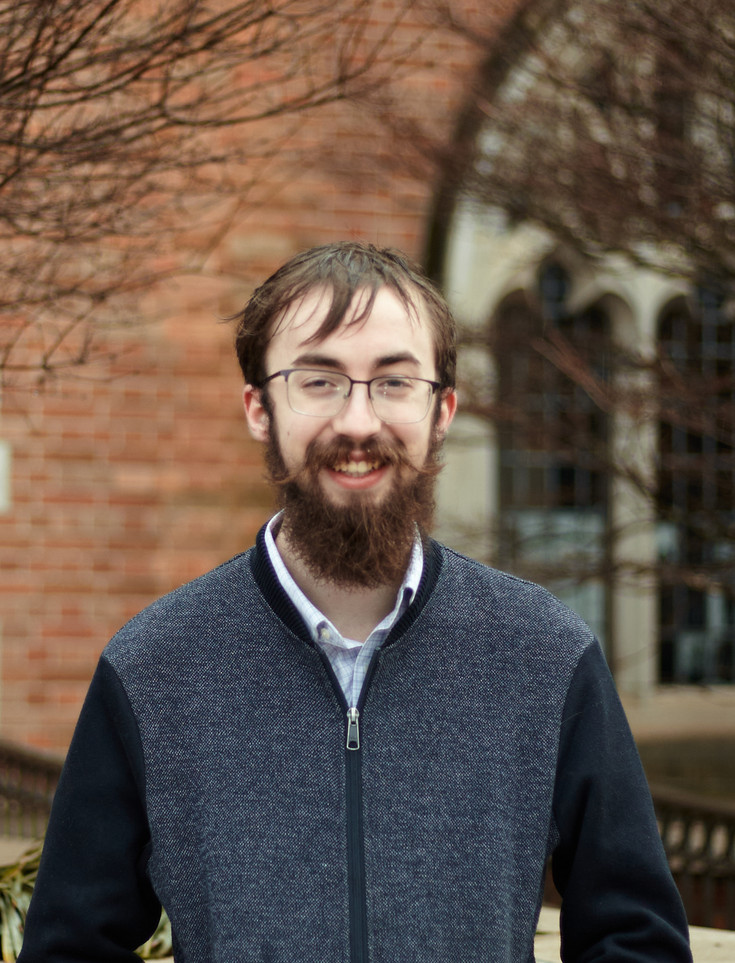}}]{Robert Wallace} is currently working toward the doctoral degree with the University of Notre Dame, advised by Dr. Collin McMillan. His research interests are in software engineering, natural language proccessing, and human attention.
\end{IEEEbiography}

\vspace{-2cm}

\begin{IEEEbiography}[{\includegraphics[width=1in,height=1.25in,clip,keepaspectratio]{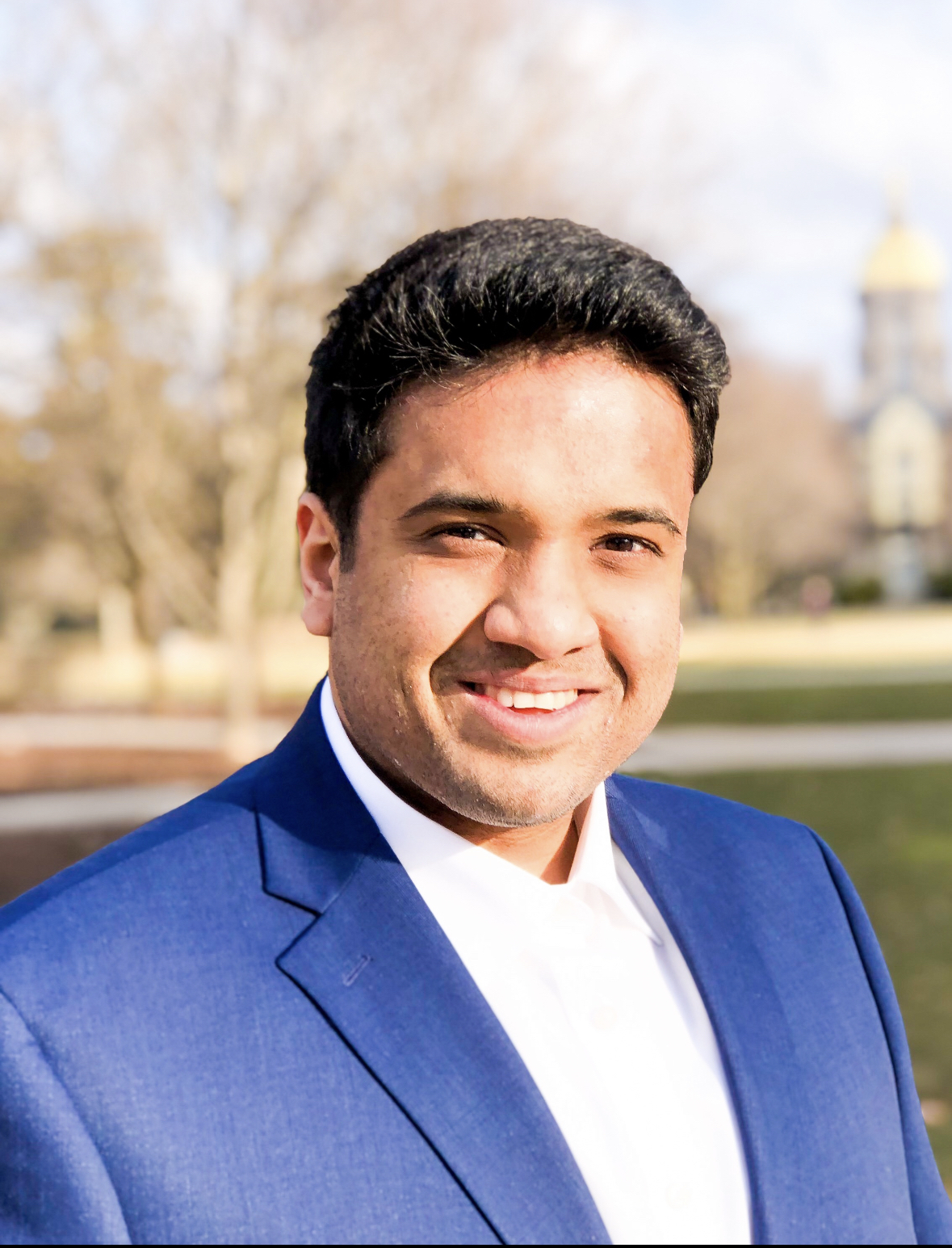}}]{Aakash Bansal} is an assistant professor in the division of Computer Science and Engineering at Louisiana State University. He joined LSU in 2024, after earning his Ph.D. at University of Notre Dame. His research is broadly in the field of AI for software engineering. Specifically, his work focuses on designing AI models of code that leverage context and human-factors.
\end{IEEEbiography}

\vspace{-2cm}

\begin{IEEEbiography}[{\includegraphics[width=1in,height=1.25in,clip,keepaspectratio]{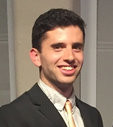}}]{Zachary Karas} is a doctoral student with Vanderbilt University, advised by Dr. Yu Huang. He studies the cognitive factors of software engineering to improve computer science education and software development. Using neuroimaging and eye-tracking, his research examines the cognitive processes underlying code writing, summarization, and comprehension.
\end{IEEEbiography}

\begin{IEEEbiography}[{\includegraphics[width=1in,height=1.25in,clip,keepaspectratio]{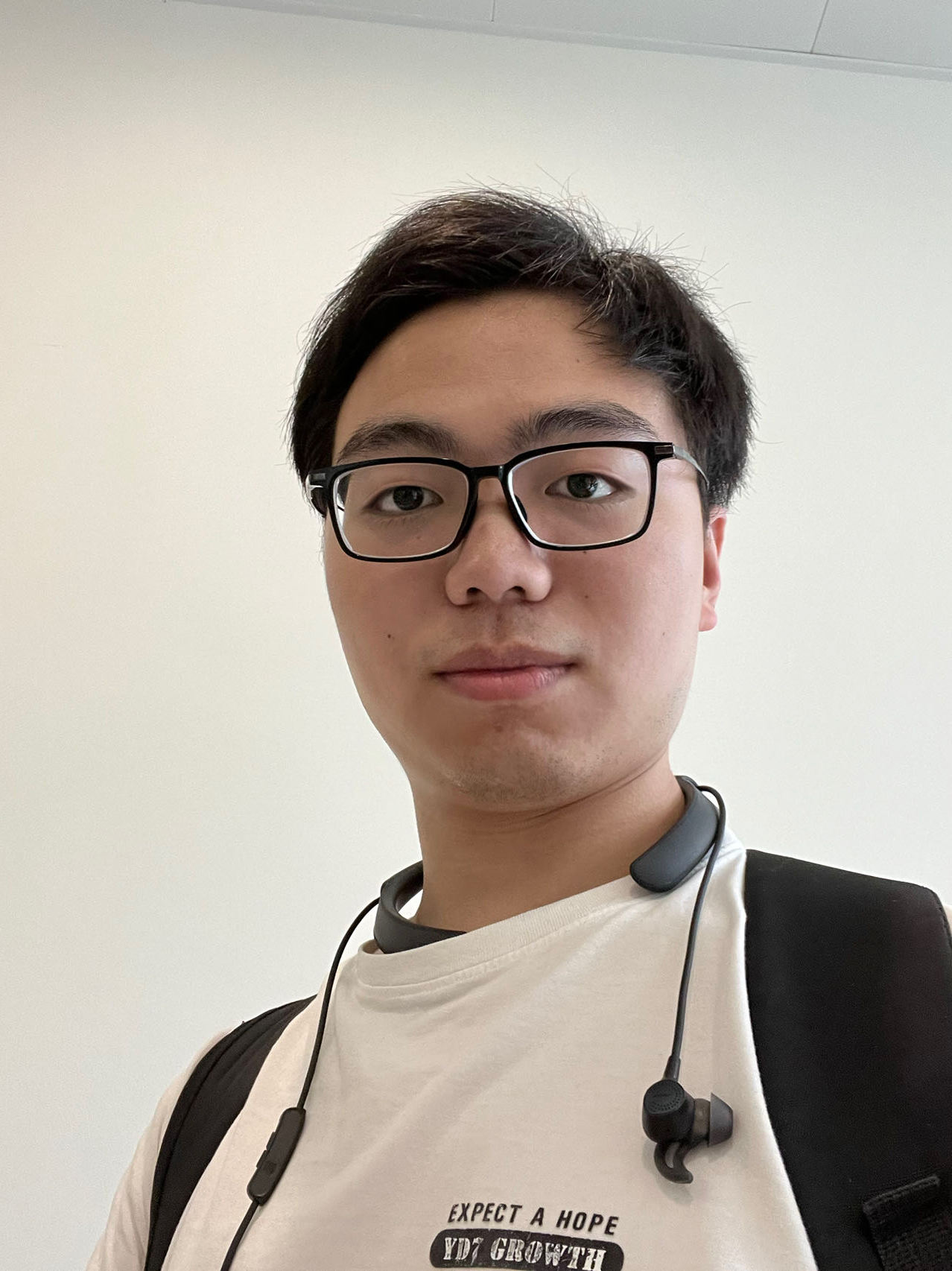}}]{Ningzhi Tang} is a Ph.D. student in the Department of Computer Science and Engineering at the University of Notre Dame, advised by Prof. Toby Jia-Jun Li in the SaNDwich Lab. His research focuses on developing human-centered tools to enhance developers' programming experience, particularly in code comprehension and debugging, while also exploring the challenges and opportunities of AI in software engineering.
\end{IEEEbiography}

\begin{IEEEbiography}[{\includegraphics[width=1in,height=1.25in,clip,keepaspectratio]{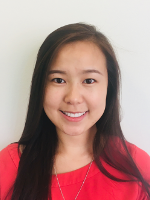}}]{Yu Huang} is currently an assistant professor in the Department of Computer Science with Vanderbilt University’s Institute for Software Integrated Systems. Her expertise lies in the fields of human factors and software engineering, with a specific emphasis on leveraging diverse modalities to explore user cognition and behaviors, and their intricate connection to AI for software engineering, sustainability for open source software, and computer science education.\end{IEEEbiography}

\begin{IEEEbiography}[{\includegraphics[width=1in,height=1.25in,clip,keepaspectratio]{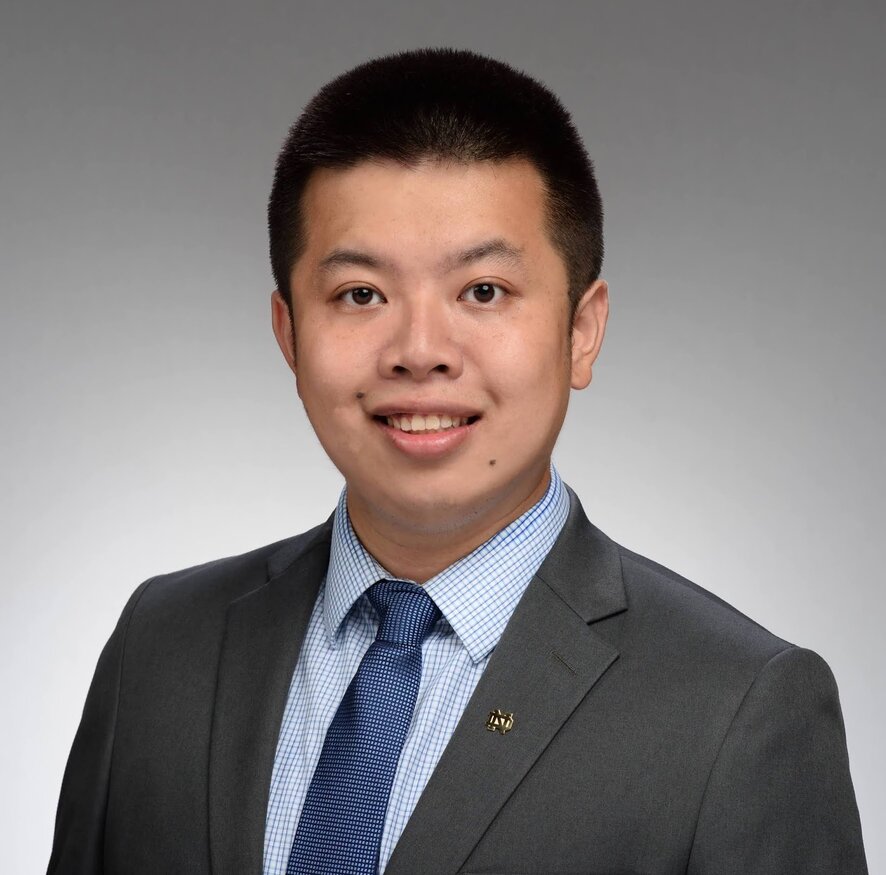}}]{Toby Jia-Jun Li} is an Assistant Professor in the Department of Computer Science and Engineering at the University of Notre Dame, where he leads the SaNDwich Lab. He also serves as the Director of the Human-Centered Responsible AI Lab in the Lucy Family Institute for Data \& Society and a Faculty Fellow at the Institute for Educational Initiatives (IEI). Toby received his Ph.D. in Human-Computer Interaction at Carnegie Mellon University, where he was advised by Brad A. Myers. 
\end{IEEEbiography}

\begin{IEEEbiography}[{\includegraphics[width=1in,height=1.25in,clip,keepaspectratio]{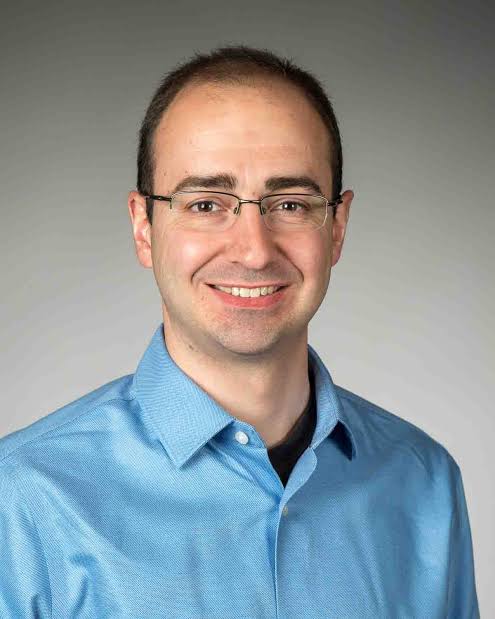}}]{Collin McMillan} is an associate professor in the
Department of Computer Science and Engineering at
the University of Notre Dame. He started with Notre
Dame in 2012 after finishing the PhD degree with the
College of William \& Mary. Dr. McMillan’s work has
since been recognized with the NSF CAREER award
and the ASEE Illinois/Indiana Teacher of the Year
award.
\end{IEEEbiography}

\end{document}